# Early stage phase separation of AlCoCr$_{0.75}$Cu$_{0.5}$FeNi high-entropy powder at the nanoscale

**Nicolas J. Peter[a], Maria J. Duarte[a], Christian H. Liebscher[a,*], Vikas C. Srivastava[c], Volker Uhlenwinkel[b], Eric A. Jägle[a,*], Gerhard Dehm[a]**

[a] Max-Planck-Institut für Eisenforschung GmbH, 40237 Düsseldorf, Germany
[b] Leibniz-Institut für Werkstofforientierte Technologien, 28359 Bremen, Germany
[c] CSIR-National Metallurgical Laboratory, Jamshedpur 831007, India

*Corresponding authors: jaegle@mpie.de, liebscher@mpie.de

| | |
|---|---|
| NJP: | peter@mpie.de |
| MJD: | duarte@mpie.de |
| CHL: | liebscher@mpie.de |
| VCS: | vcsrivas@nmlindia.org |
| VU: | uhlenwinkel@iwt.de |
| EAJ: | jaegle@mpie.de |
| GD: | dehm@mpie.de |

# Abstract

High entropy alloys are generally considered to be single phase material. This state is, however, typically a non-equilibrium state after fabrication at high cooling rates. Phase constitution after fabrication or heat treatment is mostly known for isothermal annealing only and for casts as well as rapidly quenched alloys. Knowledge on early phase separation stages of high entropy alloys and their mechanisms are missing so far. Here, we present results on phase separation at intermediate cooling rates, by characterization of gas atomized powder of the $AlCoCr_{0.75}Cu_{0.5}FeNi$ alloy. Although investigation by X-ray diffraction and Electron Backscatter Diffraction indicates a single-phase nature of the powder particles, aberration-corrected scanning transmission electron microscopy and atom probe tomography reveal a nanoscale phase separation into Ni-Al-rich B2 and Fe-Cr-rich A2 regions as well as a high number density of $3.1 \times 10^{24}$ Cu-rich clusters per $m^3$ in the B2 matrix. The observed phase separation and cluster formation are linked to spinodal decomposition and nucleation processes, respectively. The study highlights that adequate characterization techniques need to be chosen when making statements about phase stability and structural evolution in compositionally complex alloys.

# Keywords

# I. INTRODUCTION

Near-equiatomic, multi-principal component alloys – commonly termed compositionally complex alloys (CCAs) or high-entropy alloys (HEAs) – were conceptually introduced about a decade ago [1–3]. The retention of extraordinarily high compressive yield strengths of nano-crystalline HEAs up to 600°C [4], is one of the examples of promising properties possible in this new class of materials. It is believed that high configurational molar entropy favors a single ideal solid solution phase in HEAs over intermetallic compounds. This is the so-called high entropy effect, as introduced by Yeh et al. in 2004 [2]. The Hume-Rothery rules provide a guideline for the formation of a single-phase substitutional solid solution, i.e. similar atomic radii, crystal structures, electro-negativities and valences. Consequently, combinations of 3d transition metals seem to provide a suitable playground for exploration of such single-phase HEAs (among other HEA classes such as refractory metal-based alloys). This opened up new avenues, breaking the traditional concepts, and a change in paradigm of alloy design. Especially alloys based on Al, Cr, Mn, Fe, Co, Ni and Cu have been studied thoroughly, starting with the very first single-phase HEA studied by Cantor et al., i.e. CrMnFeCoNi [3]. However, a multitude of alloy compositions, satisfying the Hume-Rothery rules and predicted to show the high-entropy effect, have shown phase constitutions ranging from (seemingly) single-phase solid solutions [5] to up to four or five different coexisting phases [6], then termed CCAs. The phase constitution of an alloy sometimes changes by a slight concentration variation of one of the alloying elements, e.g. by the Al concentration in the $Al_x$CoCrCuFeNi system [7]. This renders the establishment of comprehensive microstructure-property relationships a challenging task and requires first of all a thorough microstructural characterization, particularly at small scales.

The available literature indicates that X-ray diffraction (XRD) and scanning electron microscopy (SEM) are the main characterization methods used to explore the phase constitution of HEAs [8–10]. In some studies, however, transmission electron microscopy (TEM) was also employed to visualize, and identify, by complementary electron diffraction and energy dispersive X-ray spectroscopy (EDS), the phase distribution and composition of a material [11–13]. The observations reported in literature, about phase composition and microstructure, are often based on such experiments. The idea of spinodal phase decomposition in these alloys has been put forward, but hardly ever directly observed [7,14]; especially in early stages of the decomposition. The instrumental limitations such as the lack of sensitivity of XRD or the limited resolution of SEM as well as hidden phases in TEM diffraction patterns emphasize the need of a length-scale bridging characterization approach, including high-resolution methods like aberration-corrected TEM and 3D atom probe tomography (APT). Aberration-corrected scanning transmission electron microscopy (STEM), in particular, is beneficial as it employs atomic spatial and chemical resolution in combination with EDS or electron energy loss spectroscopy (EELS). As STEM is only capable of capturing projection information, APT extends the high chemical and partially even structural resolution in projection direction. Recently, some studies have applied these methods and demonstrated, for instance, coexistence of ordered ($L1_2$) and disordered phases (A1), which were not revealed by traditional methods [15], and the existence and composition of superalloy-like channels between precipitates just few nanometers in width [16].

It is known that cooling rate strongly influences the solidification microstructure of materials by altering the solidification sequence. Singh et al. conducted a thorough series of experiments to study the kinetic differences between slow cooling during casting (10-20 K/s) and fast cooling achieved by splat quenching ($10^6$-$10^7$ K/s) [17]. The nominal composition of the HEA of this study was AlCoCrCuFeNi, similar to the alloy under investigation in the present study (AlCoCr$_{0.75}$Cu$_{0.5}$FeNi), and it was characterized by APT. Although the authors used their data to provide a model of phase evolution with cooling rate, i.e. from single phase (splat quenching) to multi-phase material (casting), the investigation of intermediate cooling rates is missing. These two processes of casting and splat quenching, with extreme kinetic restraints, do not highlight the early stages of phase separation. However, solidification at intermediate cooling rates (about $10^3$-$10^4$ K/s) might provide further insight into phase separation and support the phase evolution model. In the study by Singh et al. [17], the splat quenched samples were interpreted as single crystalline, even though dark field imaging in TEM revealed a contrast pattern indicative of changes in crystal structure. The authors did not further characterize or interpret this striking microstructural feature, as is also the case in [18].

In the present study, gas atomized AlCoCr$_{0.75}$Cu$_{0.5}$FeNi alloy powder, cooled at intermediate rates between casting and splat quenching, has been investigated. The cooling rate experienced by droplets of molten metal during gas atomization depends on droplet size. Generally, it is in the order of $10^3$ to $10^4$ K/s [19] for molten metal droplets and a diameter of ~ 100 μm, as compared to $10^7$ K/s or higher for splat-quenching [20]. Therefore, the present investigation is expected to bridge the gap between cooling rates of the experiments conducted by Singh et al. [17]. At the same time, we show by high-resolution characterization methods that the seemingly single-phase material is indeed multi-phase. The performed experiments, XRD, SEM-based techniques and conventional TEM indicate single-phase material. However, high-resolution scanning TEM (STEM) and APT are capable of revealing a nanometer-sized, interwoven phase separation of A2 and B2 phases, and the formation of a high density of Cu clusters. The observed modulated morphology and compositional variation indicates towards the early stages of spinodal decomposition. We discuss this mechanism of phase separation, in contrast to nucleation and growth of the Cu clusters. Finally, the results are compared with an existing phase prediction model to reach to a complete picture of phase evolution in compositionally complex AlCoCuCrFeNi-based alloys.

## II. EXPERIMENTAL PROCEDURES

HEA powder particles were obtained as byproduct of a spray forming technique while investigating the influence of cooling rate on the microstructural evolution in the HEA with a nominal composition of AlCoCr$_{0.75}$Cu$_{0.5}$FeNi [21]. The spray forming setup was composed of an induction melting chamber, a tundish, and a free fall atomizer. The liquid metal was heated to a temperature of 1773 K and atomized with nitrogen gas at a gas- and mass flow rate of 850 kg/h and 430 kg/h, respectively, for a pressure of 4 bars. The overspray particles not contributing to the spray-formed material on a substrate plate were collected and sieved to a particle size below 125 μm. The particle fabrication set up is shown schematically in Figure 1a). Further details can be found in [22]. Typical cooling rates for overspray particles were calculated previously to be $10^3$ to $10^4$ K/s for about 100 μm diameter particles, but depending on the particle diameter, with bigger particles experiencing a lower cooling rate [19]. Chemical analysis of the main (metal) constituents was performed via inductively coupled plasma optical emission spectrometry (ICP-OES).

XRD analysis of the HEA powder was performed using a Philips PW 1830 diffractometer equipped with a Co-K$_\alpha$ X-ray source at a wavelength of λ = 1.78897 Å. Diffractograms were acquired at a step size of 0.02°, a count time of 10 s per step and a sample rotation speed of 1 loop per second. Analysis of the acquired diffraction pattern was carried out by Rietveld analysis within the Bruker Topas software (version 5.0). Intensity was plotted against scattering vector $Q = 2\pi/d = 4\pi \sin\theta/\lambda$. High-resolution STEM and APT, showed that the alloy phase separates into A2 and B2. Using this information, the recorded XRD pattern was re-examined and lattice parameters were analyzed only for the diffractogram containing super lattice reflections (B2 phase for Rietveld analysis) and for the main peaks in the diffractogram (A2 phase for Rietveld analysis).

Thin lamellae for STEM investigations were roughly prepared in a JEOL JEM 9320 single-beam focused ion beam (FIB) at 30 kV acceleration voltage and about 2 nA current. Particles were imaged in the same machine at same conditions to show the particle shape and morphology. The lamella transfer and fine preparation were achieved in a FEI Helios 600 crossbeam machine at 30 kV down to an ion current of approximately 80 pA. Final polishing of the lamella was performed at 5 kV and around 40 pA. Channeling contrast images were acquired with the same machine at an average current of 2.8 nA, thereby revealing a qualitative orientation map. For all experiments, an Everhardt-Thornley-Detector was used in secondary electron mode.

Electron backscatter diffraction (EBSD) measurements were performed in a JEOL-6490 SEM at an acceleration voltage of 15 kV, a working distance of 30 mm and a step size of 0.5 μm on embedded and metallographically ground and polished particles. The presented results are composite images of image quality (Kikuchi pattern quality index) and either the phase map or the orientation map after filtering the acquired data set with a confidence index of 0.45 to exclude random orientations from the carbon embedding matrix. Thus, a minimum confidence index of 0.45 for the collected pixels within the HEA powder particles is inferred. The presented data was prepared using the OIM analysis software. The analysis was done three times separately, providing only A2, only B2 and both A2 and B2 as possible phases.

Scanning transmission electron micrographs (STEM) were obtained with an aberration-corrected FEI Titan Themis 60-300 microscope at semi-convergence and inner semi-collection angles of 17 mrad and 73 mrad, respectively, thus being operated under high-angle annular dark field (HAADF) conditions at an acceleration voltage of 300 kV making use of the Z-contrast characteristics of this mode, especially at atomic resolution. The same microscope was used in conventional TEM mode (parallel illumination without aberration-correction) to perform selected area electron diffraction (SAED) and to record bright field / dark field images. EDS maps were acquired using a SuperX four detector system (ca. 15 min acquisition time for each map) at different magnifications. The chemical composition and elemental spatial distribution were quantified using a principal component analysis implemented in MATLAB using k-factors obtained from the Bruker system installed at the microscope [23]. Elemental concentration integrated line profiles were extracted from these quantified EDS maps. Phase separation, observed in HAADF images, was quantitatively visualized using a MATLAB code that identifies atomic column positions (intensity peak positions) and determines the mean intensity difference to its nearest neighbors. The ordered B2 structure, with Al containing atomic columns of low intensity (high order parameter), is thereby separated from the A2 structure of homogenous atomic column intensities (low order parameter) under HAADF conditions.

To support the high spatial and hence structural resolution of the electron microscope, APT was performed. This method yields higher chemical sensitivity, but slightly lower spatial resolution than HRTEM. Six tip-shaped specimens for field evaporation were prepared from the embedded powder particles by a standard FIB-based lift-out procedure. After annular milling (sharpening of the tips), a final tip cleaning was performed with low ion energy of 3 kV. This way, the Ga-content in the analyzed volumes was kept below 0.01 at.%. The samples were analyzed in voltage pulsing mode of a Cameca local electron atom probe LEAP3000HR instrument. The representative tip shown was obtained at a set base temperature of 50 K, a pulse repetition rate of 200 kHz and with a total amount of more than 29 million ions detected at a rate of 0.01 atoms per pulse with a pulse fraction of 15 %. Traces of various impurity elements were detected during the measurement but were below 0.01 at. % and homogenously distributed in the volume, and thus were not considered for further analysis. The data reconstruction was carried out using IVAS 3.8.2 from Cameca, with a voxel size of 1 nm in each direction and isotropic delocalization set to 3 nm. Distribution analyses were conducted using different reconstructed volumes and bin sizes yielding similar results. The data presented corresponds to an entire dataset using a box size of 100 ions.

Nanoindentation was used to determine the powder's average hardness and reduced modulus in a commercial G200 Agilent standalone nanoindenter. Three quadratic grids consisting of 4 x 4 indents were measured in the cross-section of three different powder particles. The spacing between individual indents was fixed to 20 μm, while every indent's peak force was reduced by 85 % compared to the previous one to study the influence of penetration depth on the mechanical properties. The highest peak force applied, was 20 mN. The measured force-displacement curves were analyzed by the Oliver & Pharr [24] method to extract hardness and reduced modulus.

# III. EXPERIMENTAL RESULTS

Gas atomization of liquid AlCoCr$_{0.75}$Cu$_{0.5}$FeNi alloy was successfully realized to obtain spray formed consolidated deposit and overspray powder, with particle diameters of less than 125 µm, was collected from the atomization chamber (cf. schematic in Figure 1a)). Wet chemical analysis of the overspray powder of the AlCoCr$_{0.75}$Cu$_{0.5}$FeNi alloy confirmed a close proximity to the nominal composition. Impurities such as carbon and oxygen were found to be well below 0.1 at. %. The results of the chemical analysis are summarized in Table 1.

The XRD diffraction pattern of the powder sample used in this study (Figure 1b, upper curve) exhibits sharp discrete peaks. The peaks showed almost perfect agreement with the ordered B2 single phase listed in the ICDD database under PDF number 00-044-1267 for Ni$_{0.42}$Al$_{0.58}$. The corresponding indices of planes are indicated, including the additional (100), (111) and (210) superlattice reflections of the B2 structure, which are weak in intensity. A previous XRD analysis of the same powder is redrawn from [21] in light blue color. This reference diffractogram was measured on the same type of gas atomized alloy utilized in the present study, but the superlattice reflection peaks from the ordered B2 phase are of very low intensity. The diffractogram intensity was plotted against the scattering vector for a better visualization of the agreement between both measurements, as the reference pattern was collected with a different type of X-ray diffraction source and step size, i.e. Cu-K$_\alpha$ type radiation and 0.5°, respectively. Thus, plotting against the scattering vector eliminates the wavelength dependency of the diffraction pattern. The comparably low intensity of the superlattice reflection peak in the reference pattern may result from a differing radiation source chosen for the measurement, i.e. Cu radiation instead of Co, which would reduce the signal-to-noise-ratio of such weak reflections. This is related to an increase in fluorescence (background intensity) from the X-ray interaction with the material's significant iron content and the diffraction pattern being shifted towards smaller scattering angles compared to a pattern recorded with Co radiation. In addition, the 25-fold bigger step size of the reference pattern may smear out such weak reflections additionally. On the other hand, the reference spectrum was collected on classified powder of distinct particle sizes, while the present study used the entire powder size distribution. This may influence the intensity because of potential differences in cooling rate. We believe, however, that the different radiation source and step size are the main reason for the minor discrepancy. Nevertheless, the diffraction patterns reveal the same crystallographic structures of the powder. In the present study a lattice constant of 2.8739 ± 0.0001 Å was determined using the ordered B2 reference mentioned above, by making use of the Rietveld method. The lattice constant of the A2 phase was re-evaluated considering the nanosized phase separation and determined to be $a_{A2}$ = 2.8734 ± 0.00014 Å, while the B2 phase's lattice constant could be determined to be $a_{B2}$ = 2.8741 ± 0.00004 Å. A vanishing effective lattice mismatch ε of 0.025 ± 0.0035 % is calculated from these values according to $\varepsilon = 100 \cdot \frac{(a_{B2} - a_{A2})}{a_{A2}}$ and regarding propagation of uncertainty.

SEM reveals a mostly spherical particle shape with a cauliflower-like surface topography. The central particle in Figure 1c) was used for FIB machining for TEM lamella extraction. Channeling contrast images of metallographically prepared (i.e. ground and polished) particle cross-sections were acquired by FIB (Figure 1d), indicating a polycrystalline microstructural arrangement inside the particles.

Closer investigation by EBSD revealed a B2 single phase material in the phase map (Figure 2a, left) with grains being randomly orientated within the particles, as shown in the orientation map (Figure 2a, right). The grain morphology resembles a solidification structure with small equiaxed grains (grains of the chill zone) at the surface of the particle, followed by elongated as well as large equiaxed grains towards the interior of the particle. Although this structure was observed in many powder particles, especially the smaller ones contained only a few grains. Grain diameter analysis by EBSD of a powder particle composed of the polycrystalline solidification structure revealed an average grain diameter of 4.06 μm ± 2.51 μm, for 825 analyzed grains (Figure 2b). The analyzed particle has a Feret diameter of 160 μm. Its circularity and aspect ratio were determined to be 0.91 and 1.3, respectively.

SAED in conventional TEM mode was used on the extracted TEM lamella and the results are shown in Figure 3. From the single crystalline grain interior both the fundamental A2 reflections (high intensity) and B2 superlattice reflections (low intensity) can be identified in the [001] zone axis diffraction pattern (Figure 3a). Bright field imaging did not show any particular structural arrangement, as shown in Figure 3b. However, dark field imaging of the exact same location, with the objective aperture centered around a {001}-type superlattice reflection, revealed a meander-like arrangement of bright and dark contrast with a characteristic domain size of 5.78±3.81 nm for the bright regions, as illustrated in Figure 3c.

Aberration corrected STEM analysis was performed on the same lamella. A stitched HAADF overview image is presented in Figure 4a. Mostly homogenous contrast was found within individual grains of the lamella, although grain boundaries with higher intensity could be identified indicating a different composition considering the Z-contrast dependence of this imaging mode. An EDS line scan, with a step size of 13 nm, was performed from the surface to 6.5 μm towards the center of the particle, as shown in Figure 4b. The line scans show no composition fluctuations on this length scale. In contrast to the nominal composition that was confirmed on average for many particles by wet chemistry, this particular position contained slightly more Co but significantly reduced Al, resulting in an approximate overall composition of $Al_{0.6}CoCr_{0.6}Cu_{0.4}FeNi$. This composition is likely affected by overlapping K-edges used for quantification and, therefore, APT is expected to provide much more accurate absolute values. Thus, in terms of absolute composition we prefer to rely on APT data, while relative composition changes can be evaluated from EDS profiles. For further investigation, one grain boundary triple line was chosen from the overview image in Figure 4a. A precipitate phase of higher intensity was observed at the triple line of three grains as well as along the grain boundary. EDS maps obtained after using a principal component analysis of this precipitate are shown in Figures 4 c-e. The maps indicate a precipitate phase of up to ~29 at. % Ni and ~45 at. % Cu, as shown in Figures 4d and 4e, respectively. All other elements are in low concentrations, which is seen as depletion in the qualitative EDS map for Al, Co, Cr and Fe (Figure 4c).

Closer inspection of the chemical composition and distribution within a single grain was performed at increased magnification of almost two orders of magnitude by aberration corrected STEM-EDS (Figure 5). HAADF micrograph and corresponding EDS map of the same location are given in Figure 5a and b, respectively. The brighter and darker areas in the HAADF image correspond to areas of higher Fe, Co and Cr concentrations (red color) or higher Ni, Cu and Al concentrations (blue color) in the EDS map, respectively, acquired at step sizes well below 1 nm. The position of the extracted line profiles is highlighted as a white line in the EDS elemental map. The quantified line spectra of each elemental species are presented in Figure 5d. They clearly show a repetitive oscillation of areas with high and low concentrations, i.e. regions with high Fe, Co and Cr concentrations show depletion of Ni, Cu and Cr. This indicates a nanoscale phase separation of the respective elements within the grain interior.

Atomic structure and the chemical phase separation were correlated at atomic resolution using HAADF STEM and EDS mapping, as shown in Figure 6. Two magnifications were considered for the structural analysis based on the ordering of A2/B2 structures. An order parameter was defined as atomic column intensity difference of neighboring columns. In the ordered regions, this order parameter is high, due to reduced intensity of Al containing columns. In contrast, the disordered phase inherits a low order parameter as neighboring columns are of approximately same intensity. The differences of column intensities for the two phases can clearly be seen in the atomic resolution HAADF micrograph provided in Figure 6e. The interface between the A2 regions, with homogeneous column intensity, and B2 domains with alternating bright and dark atomic columns, is perfectly coherent. This is also corroborated by the micrograph's Fast Fourier Transform (FFT) shown in Figure 6h. The corresponding order parameter map is shown in Figure 6f capturing the areas of (dis)order via color-coding. The combination of micrograph and order parameter map, as shown in Figure 6g, unambiguously captures the good fit between both images. The same analysis was performed on a micrograph in which individual columns could just be resolved but are hardly visible by looking at it in Figure 6a. Calculation of the order parameter map reveals distinct and spatially separated areas of high and low (down to none) order, as seen in Figure 6b. As atomic columns are not visible for correlation, an EDS map of the same location was acquired at a pixel size below 1 nm (Figure 6d), which shows the same spatial distribution of phases as indicated by the order parameter map, i.e. Al-Ni-Cu are present in blue color, while Fe-Cr is colored in red. Domain sizes of about 5 nm each fit to the previously measured size from dark field TEM measurements.

Three-dimensional chemical information of the particle's elemental distribution at highest resolution was obtained from APT reconstructions (Figure 7). A representative reconstructed volume consisting of all atomic species and their positions is presented in Figure 7a. A separation with distinct chemical composition is readily visible with a characteristic domain size of 8.65± 0.32 nm, obtained from extracted intensity profiles. The interwoven domains caused by the chemical separation are vividly demonstrated by the iso-concentration surfaces in Figure 7b. The cyan iso-concentration surface delineates volumes with Al+Ni>30 at. %. These volumes correspond to the by electron microscopy identified ordered B2 regions, while the volumes in purple (bound by 49 at. % Fe+Cr iso-concentration surfaces) consist of the disordered A2 phase. Overall, a connected meandering structure is seen similar to the projections imaged by HAADF-STEM. A closer inspection of the Cu distribution reveals a high number density of Cu clusters only a few nanometers in size, visualized by an iso-surface of Cu>20 at. % (Figure 7c). A thorough cluster analysis, as described in [25], was performed for the Cu clusters and is illustrated in Figure 7c. A total of 2316 clusters was found, which translates into a high number density of $3.1 \times 10^{24}$ clusters/$m^3$ in the entire reconstructed volume of 745 $nm^3$. Considering that the clusters were only found in the ordered B2 phase, the cluster number density in this phase would even be higher. 5 nm thick slices through the tip's volume, from the position highlighted in Figure 7a, were extracted to visualize the elemental distribution and partitioning of Al-Ni-Cu and Fe-Cr-Co, with Co having only a slight tendency to migrate to the Fe-Cr region. Evaluation of the phase fractions can be achieved in several ways from the available data. First, a thin cross-sectional volume is cut out from an APT reconstruction's volume and the area fraction of the two phases in a two-dimensional projection is determined. Following this procedure, the thin disc depicting Cr atoms in Figure 7d was analyzed. Under the assumption, that the A2 phase is Cr-rich and the B2 phase depleted in Cr, as suggested by STEM-EDS, it was found that the A2 / B2 phases share the volume by 30 % and 70 %, respectively. To validate this result, the volume fraction of individual phases with respect to the total volume's composition can be analyzed as second measure. The average compositions of the A2 and B2 phase as well as the cluster composition, as determined by APT, are provided in Figure 8. To determine the composition of the respective phases, individual sub-volumes were extracted by creating interfaces enclosing a single phase with homogeneous composition (i. e. from the plateau region of a proximity histogram [26]). The mass spectrum corresponding to each sub-volume was analyzed independently by peak decomposition using the natural isotope abundances and the composition was averaged over all the considered sub-volumes.

It is apparent that the cluster composition reveals an elemental distribution similar to the B2 region, with just over 60 at. % Cu compared to only 9.5 at. % for the matrix phase. Investigation of the bulk, A2 and B2 compositions (Figure 8) also clarifies that for all major elements, the bulk composition $c_{Bulk}$ can only be achieved by a weighted phase contribution following approximately $c_{Bulk} = (0.65 \times c_{B2}) + (0.35 \times c_{A2})$. Additionally, the STEM-based order parameter plot (Figure 6b) can be binarized and, again, an area fraction can be obtained from that, which agrees very well with A2 / B2 sharing 35 % and 65 % of the volume, respectively. A final quantification of the volume fractions of the Al-Ni-Cu and Fe-Cr-Co constituents yielded 31.9 at.% and 68.3 at.%, respectively, by determining the points of inflection in the volume fraction plot as a function of the iso-surface concentration for Ni-Al-Cu and all Fe-Cr-Co regions, as shown in Figure 9a. The volume fraction was obtained here by dividing the volume enclosed by the respective iso-concentration surface (Ni+Al+Cu or Fe+Cr+Co atoms) over the total reconstructed volume. The corresponding inflection points would correspond to the interface between the two domains. All these measurement routes led to consistently similar results compared to the last approach, which is likely the most accurate due to its general approach of considering all ions.

Frequency analysis of each individual element was performed over the entire reconstructed volume and the results for Co (only slight partitioning tendency), Ni (representative for the B2 phase) and Fe (representative for the A2 phase) are presented in Figure 9b – d, respectively. These are compared with the ideal single phase binomial distribution and, additionally, analyzed by the Langer-Bar-on-Miller (LBM) model [27], which is generally used for decomposed materials. Co almost follows the binomial distribution; however, the LBM method reveals a rather bimodal distribution of 22 at. % and 18 at. % peak concentrations, respectively. Ni and Fe reveal a much more obvious separation with a Ni-rich and Fe-rich phases, closely following the LBM model. For Cu, this approach does not hold because of the creation of Cu-rich clusters as additional phase inside the A2 phase.

A concentration profile along the long axis of a cylindrical volume (diameter of 5 nm, length 15 nm) crossing an A2 region is presented in Figure 10. In the inset, only Al (cyan) and Cr (grey) atoms are shown along with the 49 at.% Fe+Cr iso-concentration surface. The analyzed ligament was oriented close to edge-on in order to reduce concentration errors during integration. In the ordered B2 domains Cr is almost depleted and Fe strongly reduced, while Al and Ni saturate below 30 at.%. Cu is exclusively present in ordered B2 regions at about 10 at.% (no Cu cluster was included in the analyzed volume) and only a small tendency of Co partitioning was found with about 22 at.% in the non-ordered area, and about 18 at.% with respect to the B2 region.

Finally, the particle hardness and reduced modulus were measured to be 7.42 ± 0.12 GPa and 166.6 ± 8.5 GPa, respectively. The values of individual measurements did not show a significant deviation for increasing penetration depths, which is reflected in the very small standard deviation values as well. It has to be stressed, that the edge length of the remaining permanent indents ranged from about 750 nm to 2.5 μm for maximum penetration depths of about 100 nm to 375 nm, respectively, and thus only an average value of both phases is measured. No cracking was observed around the indents.

# IV. DISCUSSION

The results of the present investigation show an early stage decomposition in the AlCoCr0.75Cu0.5FeNi alloy, belonging to the relatively new class of HEAs or CCAs. This highlights that utmost care is required while claiming that a material is single-phase, particularly when investigations are performed only by non-high-resolution techniques. The seemingly single-phase particles investigated in this study, turned out to be composed of not only two main interwoven phases, but even four phases if Cu clusters and the phases segregated at the grain boundary are taken into consideration. This multiphase arrangement could have easily been overlooked by conventional methods, especially in absence of prior knowledge of the nanoscale phase separation.

In recent years, several factors have been proposed to influence phase formation in multicomponent HEAs. They are: change in configurational entropy, effective atomic size difference, electro-negativity and valence electron concentration. The effect of these factors is often manifested in phase diagrams. These factors are very similar to the Hume-Rothery rules and were initially found to identify solid solution forming ability, dividing the experimentally observed data into regions of mainly solid solutions, intermetallics and bulk metallic glasses [27, 28]. However, the investigated alloy would fall into a region of such phase diagrams, which cannot unambiguously be related to solid solution or intermetallics former. These rules do not consider kinetics, which might be a major contributor to the phase separation resulting in such ambiguities. Thus, the investigated powder is a suitable example to underline the need for both a better phase prediction model in this new class of alloys and that only adequate microstructural characterization is capable of eliminating the ambiguity.

XRD was applied to the successfully produced HEA powder with only B2-type reflections observed in the diffractogram (Figure 1). This apparently single phase structure matches well with XRD diffractograms obtained by Ivchenko et al. who showed, for an ultrarapidly quenched HEA (cooling rate of $10^6 – 10^7$ K/s) of similar nominal composition in the as-quenched state, an apparently single-phase B2 type microstructure [18]. However, making use of the Rietveld analysis to make quantitative phase composition statements is especially difficult in such phase separated multicomponent alloys. The complexity results partially from the overlapping major diffraction peaks. Further indication would give the site occupancy factor. But the multicomponent alloy contains six elements to be distributed over only one or two relevant unit cell sites, which introduces a high degree of uncertainty. As a consequence, only a weak indication remains based on the relative peak height of the diffractogram intensity. This may be difficult to judge as major peaks of the A2 and B2 phase overlap, but without having knowledge about how much overlap to really expect. However, simulation of diffractograms and comparison to experimentally obtained ones may be an alternative route to obtain more insights in the actual phase content. Still this procedure would require knowledge about the fine phase separation, which would require further experimental work. Based on this, additional EBSD analysis on the powder particles produced a phase map that identified all Kikuchi patterns of a particle as ordered B2 phase (Figure 2). However, running a phase analysis on the same dataset with different input crystal structures, i.e. only A2 or both A2 and B2, results in a single A2 phase particle or a completely random phase distribution – almost on a pixel base, respectively. This is a consequence of the method's phase identification procedure of rather recognizing the geometry of a Kikuchi pattern than the presence of superlattice reflections. Consequently, discrimination between A2 and B2 is almost impossible. These ambiguities should be remembered when trying to identify single phase materials, especially in the HEAs that are known to form finely dispersed microstructures at the nanoscale, which push the resolution capabilities of the methods to their limits.

Conventional TEM investigations in terms of SAED and bright field / dark field imaging increase the resolution of the phase identification and enable direct visualization of the microstructure, in contrast for instance to XRD. Although SAED and bright field imaging do not indicate phase separation, dark field imaging with the objective aperture centered around a superlattice reflection provides first hints of separation. Consequently, re-evaluation of the apparently B2 diffraction pattern then appears to be an almost perfect combination of a coherent network of areas including chemical ordering (B2) and areas without (A2). The small lattice mismatch of only 0.025 %, found by XRD, agrees well with a picture of highly coherent and coexisting domains. Care has to be taken with a lower resolution limit for dark field imaging originating from the aperture diameter and lens aberrations. Ivchenko et al. [18] as well as Singh et al. [17] also observed contrast differences in their quenched samples by dark-field imaging, but noted that EDS on their quenched specimens showed near-uniform distribution of the constituent elements and were therefore regarded as single phase. Consequently, the assumption of a single-phase microstructure in the as-quenched state is not unambiguously shown. Therefore, it is concluded that TEM in combination with electron diffraction should at least be performed for phase identification. However, the obtained results from such methods may only provide hints for phase separation. Especially for finely dispersed and chemically ordered phases, dark field imaging is recommended at suitable magnifications. Similar to Ivchenko's and Singh's studies, no elemental concentration fluctuations are found in the present study, thus pointing towards a single-phase material (Figure 4), although a decomposition pattern was observed in dark field TEM imaging. Only by increasing magnification, thereby reducing pixel size, is this phenomenon resolved (Figure 5). Consequently, high-resolution methods like aberration-corrected STEM and APT are necessary to make a definite statement on the material's phase constitution. The combination of both methods allows for the highest structural and chemical resolution for phase identification [30,31]. The splat-quenched samples from Singh et al. [17], described as single phase (AlCoCrCuFeNi) polycrystalline material with a non-uniform chemical ordering, appear in TEM similar to the powder particles investigated in the present study (AlCoCr$_{0.75}$Cu$_{0.5}$FeNi). However, STEM and APT measurements in the present investigation clearly reveal a chemical decomposition of the constituent elements (Figures 5-7), resulting in separation of an Al-Ni-Cu-Co-rich ordered B2 crystal structure and a Fe-Cr-Co-rich disordered A2 crystal structure (Figures 7, 8 & 10), in addition to almost pure Cu cluster formation in the ordered phase (Figure 7c). According to the ternary Fe-Cr-Co phase diagram, the revealed A2 phase falls inside the alpha region, which would likely cause ordering. However, it was also shown before that addition of Cr to FeCo-based alloys lowers the order-disorder (B2 ↔ A2) temperature by about 5°C per atomic percent of Cr [32]. In the present case, the disordered A2 phase contains about 25 at. % Cr, which would significantly lower the transition temperature and therefore promote a disordered phase [33]. In addition, work on the kinetics and mechanisms of chemical ordering suggests that increasing the Cr content may have a significant effect on the kinetics of ordering in such alloys besides lowering the transition temperature [34].

Besides being a suitable model system to disclose the importance of a scale-bridging phase analysis, the particles obtained from the gas atomization process reveal further insights into the phase evolution of this Al containing 3*d* transition metal type HEA or CCA. The often mentioned spinodal decomposition in such alloys [2,14,17,35,36], to explain the evolution of certain microstructural arrangements, could hardly ever be proven and was not yet directly observed in experiment. However, few studies discuss the precipitation of phases during thermal annealing [18]. Microstructures in many multi-phase CCAs show a maze-like appearance [35], but some appear akin to the Ni-base superalloy structure with cuboidal precipitates in a matrix [16]. Since the Ni-based superalloys decompose into γ / γ' two-phase structure by homogenous nucleation [37], more research should be undertaken to clarify the decomposition mechanisms in the relatively new class of HEAs or CCAs, i.e. if nucleation or spinodal decomposition governs the process. In this regard, the investigated powder particles in the present study revealed a presumably early stage of alloy decomposition, achieved by the intermediate cooling rate of the gas atomization process. The interwoven, meander-like pattern throughout the sample, identified by dark-field imaging (Figure 4c) and APT, in combination with the atomic-resolution STEM structural observations of a coherent A2 / B2 phase separation (Figure 7), respectively, indeed indicate a spinodal decomposition process, as sketched in [38] for AlNiCo-based magnets. The small wavelength of a few nanometers only might be related to the low diffusivity of constituent elements of HEAs in their respective alloy matrix, often referred to as sluggish diffusion [39]. In addition, frequency distribution histograms of individual elements, calculated from the APT data, show the presence of compositional modulations in Figure 9. To compare, a binomial distribution centered on the bulk elemental concentration shows the deviation of the experimental data with respect to a homogeneous composition. The experimental data was fitted according to the LBM model [27], which is commonly used on the analysis of phase separation during spinodal decomposition [40]. The LBM model can be explained by the sum of two displaced Gaussian distributions centered on the composition of the studied element in the respective phases. As observed in Figure 9, the very good agreement of the model with the experimental data can be used as indicator of a spinodal decomposition. The LBM model has been shown to be suited especially for analyzing early stages of this decomposition mechanism [41]. However, as the decomposition is not only seen with respect to the chemistry but also in terms of ordering, the study rather observes a conditional spinodal, as proposed by Allen and Cahn [42], instead of a classical spinodal decomposition, which does not involve crystalline transformations [43]. In this case, the ordering is required to promote the chemical separation. The nucleation and growth of a second phase in a matrix would be restricted to self-contained areas in that matrix, in contrast to an interconnected network. However, it has to be noted that a high undercooling might lead to high nucleation rates promoting the formation of interconnected networks as well. In addition, instead of a spinodal type, one could also imagine an eutectic type decomposition. Investigation of the alloy's phase diagram would provide reasonable guidance to answer which decomposition mechanism prevails. However, so far only pseudo-binary phase diagrams of very few compositionally complex multicomponent alloys are available, rendering this approach currently hard. Judgment based purely on observation of the microstructure might leave room for ambiguities. Nevertheless, our microstructural observations provide suitable indications for why the decomposition mechanism is rather of spinodal than eutectic type. We observe a decomposition at the nanoscale, which is homogeneously distributed within individual powder particles. Such very fine-scaled microstructures are typically expected in case of a solid state reaction such as spinodal decomposition. In case of an eutectic reaction, where the liquid phase is involved and diffusion processes are much faster, usually no such nanoscaled microstrcutures are expected. In addition, it should be mentioned that the powder particles completely consist of this nanoscaled microstructure. If this microstructure would have been formed by an eutectic mechanism, that would imply that the alloy's composition is exactly at the eutectic composition. Considering the six elements present, this is very unlikely. Furthermore, that would imply that the free energy of the A2 and B2 phase should be (at least close to) identical in case of an eutectic decomposition, which is also assumed to be unlikely. Manifested by differing $T_0$ lines in the phase diagram, one phase would likely be preferred to form primary precipitates in case of an off-eutectic composition leading to rather coarse "sunflower"-like microstructures, which initially were shown to decompose through a eutectic reaction and subsequently decompose on a smaller length scale by a spinodal mechanism [44].

The formation of Cu clusters as a third microstructural feature, at high density within the B2 phase, hasn't been observed experimentally so far. The composition of these clusters was analyzed by a cluster analysis algorithm and was found to consist of over 60 at. % of Cu (Figure 8). The remaining chemistry follows the pattern of the host B2 phase (Figure 8). This could either be a measurement artifact but more likely reflects the actual composition of these clusters, as exemplified for Cu clusters in an irradiated Fe-Cu system [45]. Considering the existence of these clusters in the B2 phase, they most likely adopt the A2 crystal structure, as no particles of different crystal structure were identified in HAADF-STEM images, although mainly being composed of A1-forming Cu. This could be a compositional effect, with 60 at. % being thermodynamically not enough for a structural/phase transformation. It could also be a confinement effect, in which the strain energy would be too high for these small clusters (about 500 atoms on average) to maintain a stable A1 lattice. This would agree with the discussion of Liebscher et al. [46] that coherency stresses are the reason for A2 crystal structure adoption of Cu-rich regions in Cu-Cr films. In addition, this confinement effect would also agree with the formation of A2 type Cu precipitates below a size of 2-5 nm in a Fe-3Si-2Cu alloy, which sequentially transform from A2 over R-phases into A1 structure with increasing size [47]. The formation of these clusters within the B2 phase may be attributed to the fact that Cu has at least twice as high mixing enthalpies as for the Cu-Cr and Cu-Fe pairs of the A2 phase. But even inside the B2 phase the mixing enthalpies for all remaining pairs, i.e. Cu-Ni, Cu-Co and Cu-Al, are slightly positive or just around zero for Cu-Al (Table 2). The investigation of the Cu-Ni-Al ternary phase diagram revealed the presence of Ni-Al phases compared to Cu-rich phases [48], underlining the thermodynamic tendency of Cu to precipitate from the matrix. Consequently, Cu tries to avoid incorporation into the B2 matrix by forming independent clusters eventually leading to big Cu precipitates of very similar composition as shown for lower cooling rates or special heat treatments of similar alloys [17,18]. In contrast to the spinodal decomposition into A2/B2 interwoven compartments, the Cu clusters are considered to form by a nucleation and growth process, as previously observed in the Cu-Ti and Cu-Fe systems as well [49]. The clusters can be considered nuclei, because of their existence in a body centered cubic crystal structure. Few atoms form small clusters of high internal pressures, which eventually transform into face centered cubic phase with increasing number of atoms, thus increasing size and reducing internal stresses. Because of the very small size of Cu precipitates (consisting of a mere few hundreds of atoms each) we were only able to analyze them by cluster analysis, which does not allow to extract meaningful concentration profiles to visualize a possible sharp concentration jump between matrix and precipitate. In addition, due the limited spatial resolution of APT in general and trajectory aberrations (so-called local magnification effect) in the vicinity of phase boundaries such sharp concentration jumps are not resolvable by more conventional analysis methods like simple line profiles. Nevertheless, the small number of atoms in a cluster would lead to spherical precipitates of about $1 - 1.5$ nm in diameter at a density of about $3 \times 10^{24}$ / m$^3$. These numbers corroborate the earlier numbers obtained by Liu & Wagner in a Ni-Cu-Al ternary alloy, which provides the base of our studied material [50]. Consequently, it is concluded that a nucleation and growth mechanism prevails and the clusters belong to the nucleation or early growth stage.

In addition to Cu clusters in the B2 matrix, a Cu-Ni-rich phase forms along the grain boundaries (Figure 4). However, compared to the spinodal decomposition and the nucleation processes discussed above, the formation of this phase at the grain boundary does not belong to these classes, but rather appears during solidification of the melt. Recent Thermo-Calc calculations based on the CALPHAD approach found that upon solidification from the melt, the A2/B2 phases are formed first, but with the remaining melt being rich in Cu and Ni [21]. Eventually, this Cu/Ni-rich liquid phase is predicted to crystallize in form of a face centered cubic phase that we believe is found experimentally along the grain boundaries. Indeed, the experimentally found concentrations of about 45 at. % (Cu) and 30 at. % (Ni) at 1473 K are not too far off from experimental results reporting a mainly Cu- but also Ni-rich phase [17] in these alloys.

In the studies of Ivchenko et al. [18] and Singh et al. [17], which use a similar alloy (AlCoCrCuFeNi) compared to the present study (AlCoCr$_{0.75}$Cu$_{0.5}$FeNi), the quenching process at highest cooling rates (≥ 10$^6$ K/s) resulted in a similar kind of decomposition pattern with slightly bigger periodicity of 11.06 ±4.86 nm and 8.67 ± 4.57 nm, respectively. These values were obtained by evaluating intensity line profiles of their electron micrographs with respect to the peak-to-peak intensity separation. However, the interpretation of single phase B2 material is reported in both cases, although dark field imaging using a superlattice reflection strongly indicates phase separation. The alloy used in the present study had a slightly deviating nominal composition with the purpose of avoiding demixing or segregation of Cu due to its high mixing enthalpy with most constituent elements (Table 2) and promotion of the body centered type crystal structure formation by increasing the alloy's Cr content [51]. Consequently, the alloy is expected to be more stable in terms of decomposition. Nevertheless, for the cooling rate of the gas atomization process, phase separation occurred, as discussed above, and the performed experiments can be considered an extension of the kinetic experiments in [17]. Their performed experiments for very high (> 10$^6$ K/s for splat quenching) and low cooling rates (ca. 10 – 20 K/s for casting) draw a picture from an apparently single phase solidification to the formation of several phases for slow cooling rates. Consequently, our intermediate cooling rate (10$^3$ – 10$^4$ K/s for gas atomization) captures the early stage of the complex phase formation. This complex pattern was observed for the same alloy in [18] as well, although for annealing experiments rather than varying the cooling rate. Both these experimental studies show the evolution of phases for both a highly non-equilibrium state and the more equilibrium-like state by ultrarapid quenching and casting/annealing, respectively. However, they do not suffice to explain the early stages of phase separation and their mechanisms. The present study achieved this extension and allows for the proposition of a phase evolution model as schematically shown in Figure 11. At highest cooling rate, the formation of a polycrystalline single phase solid solution is favored over the formation of a bulk metallic glass, as for instance the effective atomic size difference is too small [28,29,52]. For moderate cooling rates, first grain boundary segregation from the melt would occur, followed by a separation of the grain interior by spinodal decomposition into regions of A2 and B2 phases at the nanoscale. For our composition and experimental conditions, a dominant B2 phase was found in terms of volume fraction. Subsequently, the onset of Cu precipitation occurs by a nucleation process within the B2 region, due to the high mixing enthalpy of Cu with most constituent elements. Further annealing from this point will lead to coarsening of the nanoscale phases as found previously. This final stage has been well studied for cast and annealed samples in [17,18] and could be considered close to equilibrium state. It is unclear, yet, if the presently found spinodal type decomposition can be suppressed at very high cooling rates and whether this arrangement will be thermally stable. So far, as the mechanical properties are concerned, the gas atomized powder particles reveal a hardness of about 7.5 GPa (ca. 750 MPa H$_v$). This hardness value is not generally achieved for rapidly quenched materials but only after annealing at 550°C. Such powder seems to provide a promising base material for additive manufacturing methods such as selective laser melting, with similar cooling rates, to obtain a pre-determined microstructure with promising mechanical properties. The question arises if the unique mechanical properties of apparently single phase high entropy alloys are indeed an intrinsic material property (intrinsic size effect), or if these alloys – especially in early studies – were decomposed at similar small length scales as the present powder particles. It turns out that only very few truly single phase high entropy alloys remain like the original Cantor alloy. Dislocations traveling through the present interwoven microstructure would constantly experience barriers of different chemical composition and crystal structure, which would likely increase their strength (as indicated by our performed nanoindentation experiments). However, lattice strain strengthening by hindering dislocation motion, as indicated in [53], seems unlikely, at least for the present alloy, since no significant incoherency could be detected by XRD and highest-resolution aberration-corrected STEM. Consequently, the search for a reliable mechanistic explanation of improved mechanical performance of the new class of multicomponent alloys remains open.

Finally, an attempt has been made to generalize our findings on phase separation by plotting the (Fe + Cr + 0.5*Co) atomic concentration against the (Ni + Al + 0.5*Co) atomic concentration, representing "Matrix former" and "Precipitate former", respectively (Figure 12a). Co was added as half the alloy's overall composition to each axis, since it has very low demixing tendency. The data points besides the present HEA are taken from [54–58]. The plot reveals a trend that the CCA can be interpreted as an extension of conventional steel alloy design in terms of B2 precipitation in an A2 matrix. In addition, if only B2 formation in the A2 matrix is considered, the precipitate volume fraction is plotted against the alloy's overall Al content (Figure 12b). It is clear that except for the maraging steel, produced by additive manufacturing and aged only by the intrinsic heat treatment (IHT) during this process, a linear relation becomes apparent. Again, the present HEA can be considered as an "extrapolation" of the microstructural states known from steels. The maraging steel produced by additive manufacturing does not follow the trend because IHT is not strong enough to induce complete precipitation. Further ageing up to peak hardness pushes this steel, too, close to the linear relation found. It seems that there is no deviation from the dependence of B2 volume fraction formed on Al content even at the high Ni and Al compositions found in HEAs. The interwoven precipitate morphology, however, presumably originating from spinodal decomposition, is distinct from steels, where individual, spherical precipitates are typically found and are interpreted as the result of a nucleation and growth transformation.

## V. CONCLUSIONS

$AlCoCr_{0.75}Cu_{0.5}FeNi$ powder was successfully produced by gas atomization. The powder was thoroughly characterized by conventional as well as high-resolution methods in terms of microstructure and phase constitution. The following conclusions are drawn from this study:

1. The experiments at an intermediate cooling rate of about $10^3$-$10^4$ K/s, inherent to gas atomization, and slightly adjusted nominal composition ($AlCoCr_{0.75}Cu_{0.5}FeNi$ compared to the equiatomic AlCoCrCuFeNi) to suppress phase separation, provided a suitable system to study early phase decomposition in this type of alloy.

2. The results point towards a phase separation by spinodal decomposition into Ni-Al-rich B2 type and Fe-Cr-rich A2 type regions at the nanoscale with similar Co content of 18 at. % and 22 at.%, respectively. In addition, Cu cluster form in the B2 region at a high number density of 3.1 x $10^{24}$ clusters/m$^3$. This early stage of phase separation can be considered as missing link to explain the phase evolution from a near equilibrium solid solution single phase. It bridges the gap between the non-equilibrium quenched state and the at least close to equilibrium state of the previously in literature described microstructures.

3. The powder particles showed high hardness of 7.5 GPa, which was previously reached only by annealing of rapidly quenched material. This indicates the importance of the high interface density within the material.

4. The analysis suggests a trend of the investigated HEA behaving like an extreme case of steels, considering the B2 occurrence and volume fraction of precipitate forming elements like Ni and Al.

# ACKNOWLEDGMENTS

The authors would like to cordially thank Benjamin Breitbach and Siyuan Zhang (Max-Planck Institut für Eisenforschung GmbH) for assistance with the performed X-ray diffraction experiments and providing the EDS principal component analysis and quantification code used. Additionally, Phillipp Kürnsteiner is cordially acknowledged for providing steel data and Baptiste Gault for discussion about the APT results. Uwe Tezins and Andreas Sturm (Max-Planck-Institut für Eisenforschung GmbH) are thanked for their support to the FIB and APT facilities. Financial support from DFG within the Priority Programme "Compositionally Complex Alloys – High Entropy Alloys (CCA-HEA)" is acknowledged under the project numbers DE 796/13-1, JA 2482/3-1, and UH 77/11-1.

# AUTHORS CONTRIBUTIONS

NJP performed most experiments, designed, coordinated this research and drafted the manuscript. MJD carried out APT experiments and data analysis. CHL created order parameter maps by digital image processing. VCS was involved in discussions about the results and worked on the manuscript. VU provided the material studied in this work and discussed about the results. EAJ coordinates the research project the presented work belongs to, consequently financed and discussed the results and worked on the manuscript. GD acted as PI, advised and provided guidance to finalize the manuscript. The authors read and approved the final manuscript.

# COMPETING INTERESTS

The authors declare that they have no competing interests.

# TABLES

Table 1: Powder composition determined by ICP-OES chemical analysis and APT.

| Element | Co | Ni | Al | Fe | Cr | Cu |
|---|---|---|---|---|---|---|
| Concentration (at. %) by ICP-OES | 19.6 | 19.5 | 19.5 | 18.8 | 12.9 | 9.57 |
| Concentration (at. %) by APT | 20.11 | 19.97 | 20.25 | 20.51 | 11.61 | 7.55 |

Table 2: Mixing enthalpies $\Delta H_{mix}$ [kJ/mol] of the constituent element pairs as reproduced from [15,21].

|  | Al | Ni | Cu | Co | Fe | Cr |
|---|---|---|---|---|---|---|
| Al | - | -22 | -1 | -19 | -11 | -10 |
| Ni |  | - | 4 | 0 | -7 | -2 |
| Cu |  |  | - | 6 | 13 | 12 |
| Co |  |  |  | - | -1 | -4 |
| Fe |  |  |  |  | - | -1 |

# FIGURES

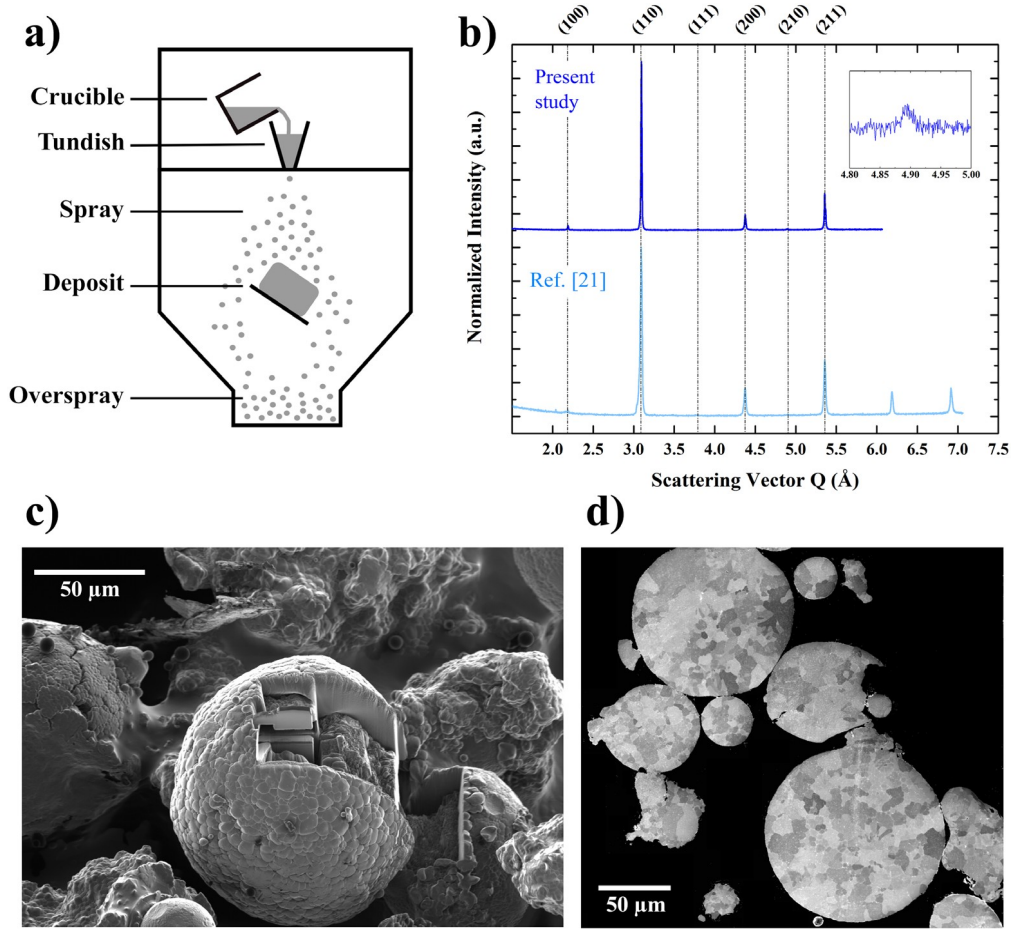

Fig. 1: The schematic design of the spray atomization chamber as introduced in [22] (a) and measured X-ray diffractogram (B2 phase reflections highlighted) for the powder used, along with a reference of the same powder analyzed in [21] (b). Surface topography of the fabricated powder particles and location of TEM lamella extraction (c) along with an ion channeling contrast image of the metallographically prepared particle cross-sections, embedded in a conductive carbon-based matrix (d).

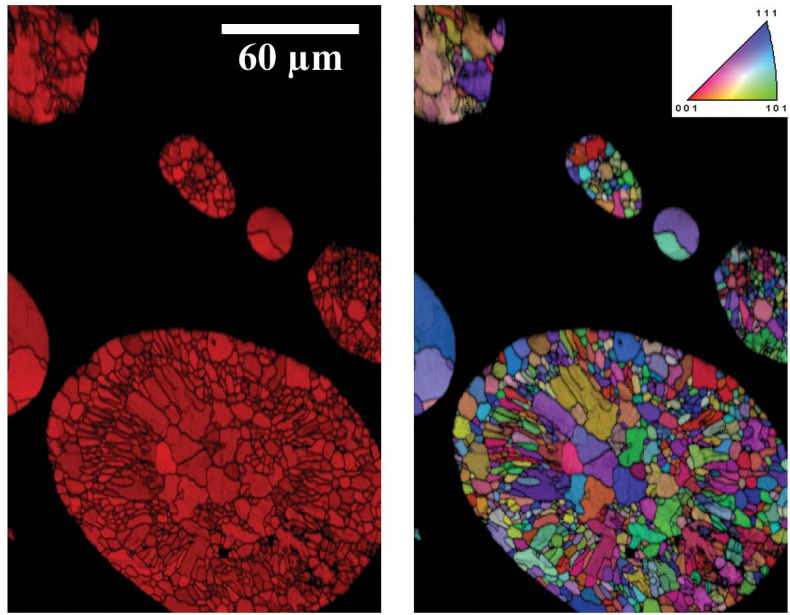

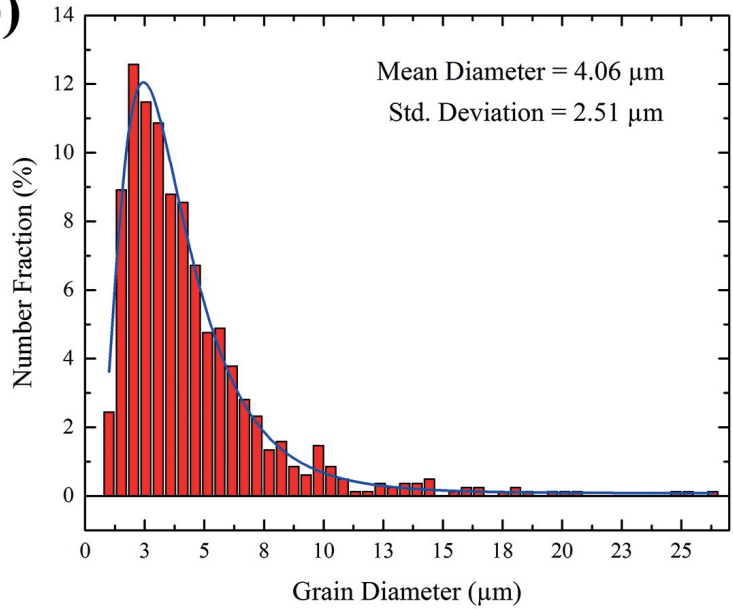

Fig. 2: Phase- and orientation maps acquired by EBSD (a) from which a grain diameter distribution (b) could be extracted (825 grains analyzed).

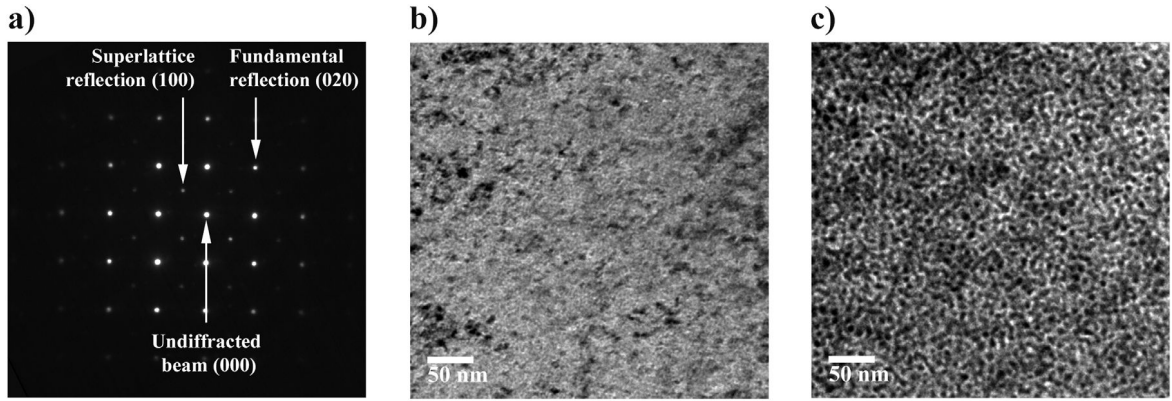

Fig. 3: Single crystalline diffraction pattern in [001] zone axis indicating B2 ordering (a). Bright field imaging in this orientation within a grain (b) does not show preferred orientations, while dark field imaging with the aperture centered around a superlattice reflection (c) reveals a meander-like structure resembling a spinodal decomposition pattern with a wavelength $\lambda$ between 5 nm and 10 nm.

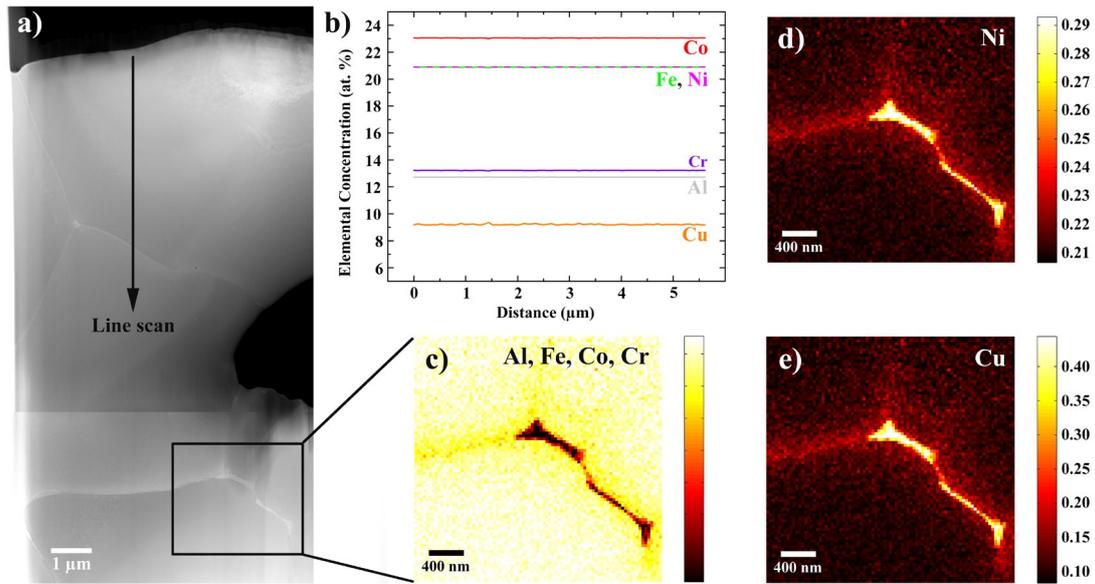

Fig. 4: Stitched HAADF-STEM overview micrograph of the complete lamella (a) used for various EDS measurements. The line scan (b) shows an apparently homogenous chemical composition. Closer inspection of a grain boundary by HAADF-STEM shows precipitation of a Cu- and Ni-rich phase (colorbar in atomic fraction) (d & e), while all other elements are almost excluded from that phase, seen in the representative map for Al, Fe, Co and Cr (qualitative colorbar of atomic fraction) (c).

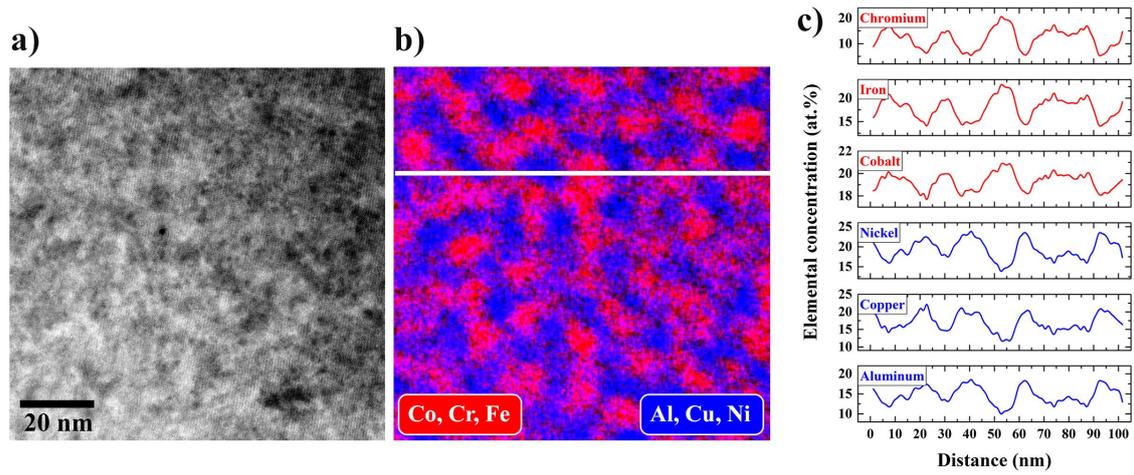

Fig. 5: Higher-resolution HAADF-STEM analysis of the grain interior (a) and associated EDS map (b). Red color is representative for Fe, Co and Cr distribution, while Ni, Cu and Al follow the blue distribution pattern. Line profiles of every element are provided in (c), extracted from the respective fully quantitative elemental map along the indicated line in (b).

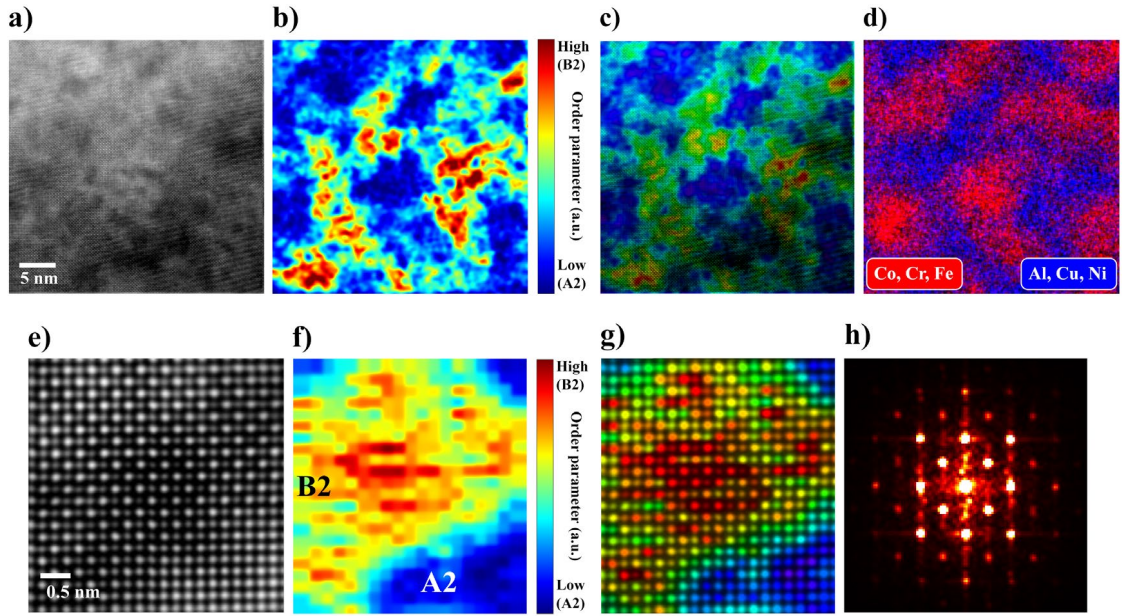

Fig. 6: High-resolution structural and chemical correlation of the HEA's grain interior. HAADF-STEM micrograph resolving two different atomic column arrangements and contrast differences that indicate chemical inhomogeneity (a). A mean nearest neighbor atomic column intensity difference (the order parameter) plot of the HAADF image indicating structural compartmentation (b). Micrograph and intensity difference plot composite image (c). Blue represents no intensity difference, while red represents highest differences as expected for A2 and B2 structures, respectively. The associated EDS map (Co, Cr, Fe in red; Al, Cu, Ni in blue) correlates the structural compartmentation to a chemical partitioning (d). Highest-resolution HAADF-STEM image at the interface of two different compartments reveals B2 and A2 crystal structures (e). The micrograph's order parameter plot reveals the differences in atomic structure at highest resolution quantitatively (f). Both the micrograph and the plot are overlaid to visualize the phase separation (g). The FFT of the micrograph reveals the same B2 superlattice type of reflections as conventional SAED (h).

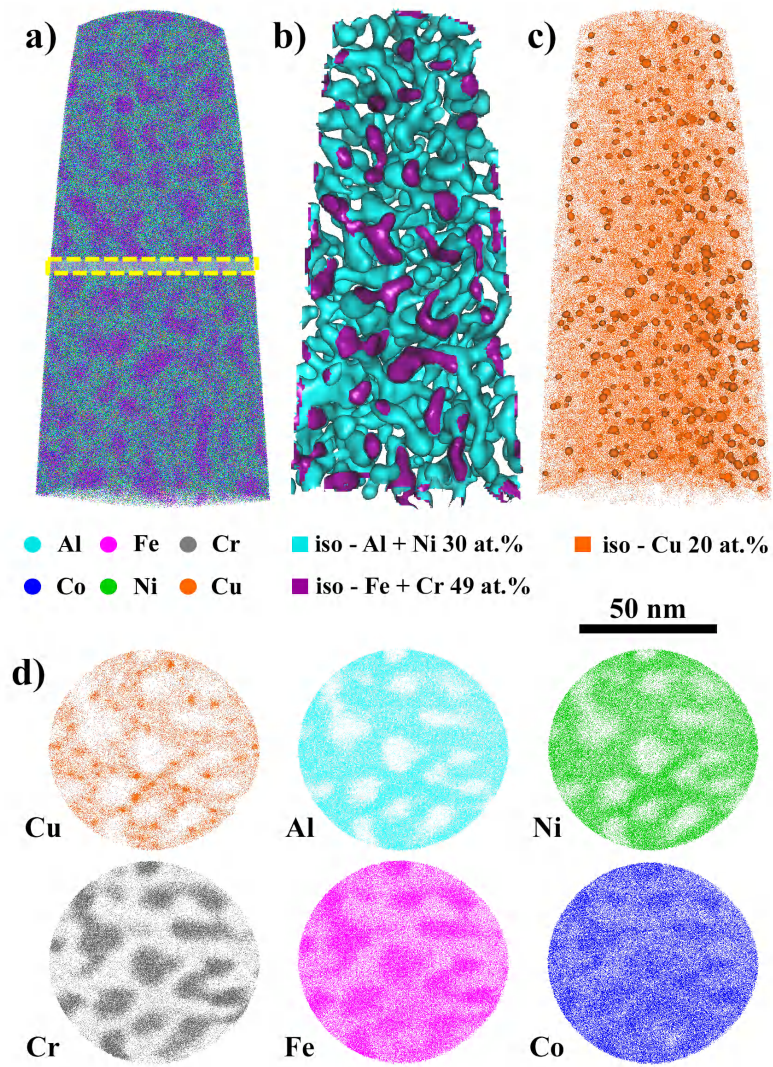

Fig. 7: General overview of the APT reconstructed volume (a). Iso-concentration surfaces representing interfaces between an Al-Ni-Cu-rich matrix (outer transparent region from the cyan iso-surface) and dendritic-like Cr-Fe-Co-rich regions (inner section highlighted in purple) (b). Overview of Cu clusters within the Al-Ni-Cu-rich region (c). Chemical partitioning shown by elemental maps of 5 nm thick cross-sections taken from the volume inside the dashed yellow line in (a) (d).

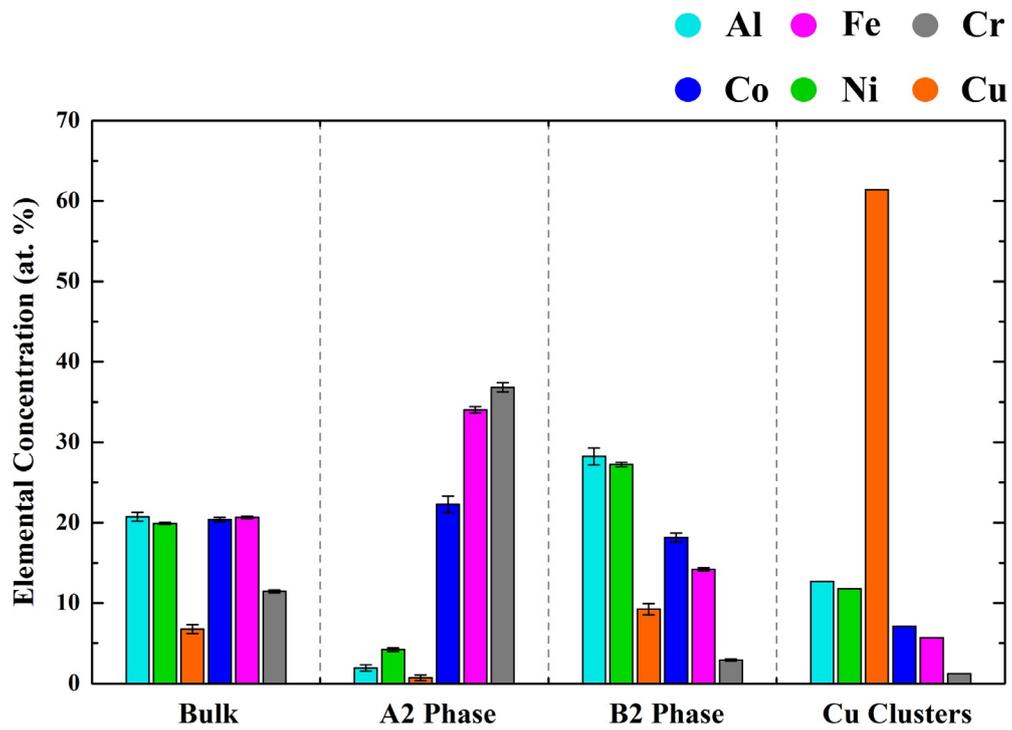

Fig. 8: Comparison of the chemical composition of the different phases present in the APT reconstructed volumes including the Cu particles.

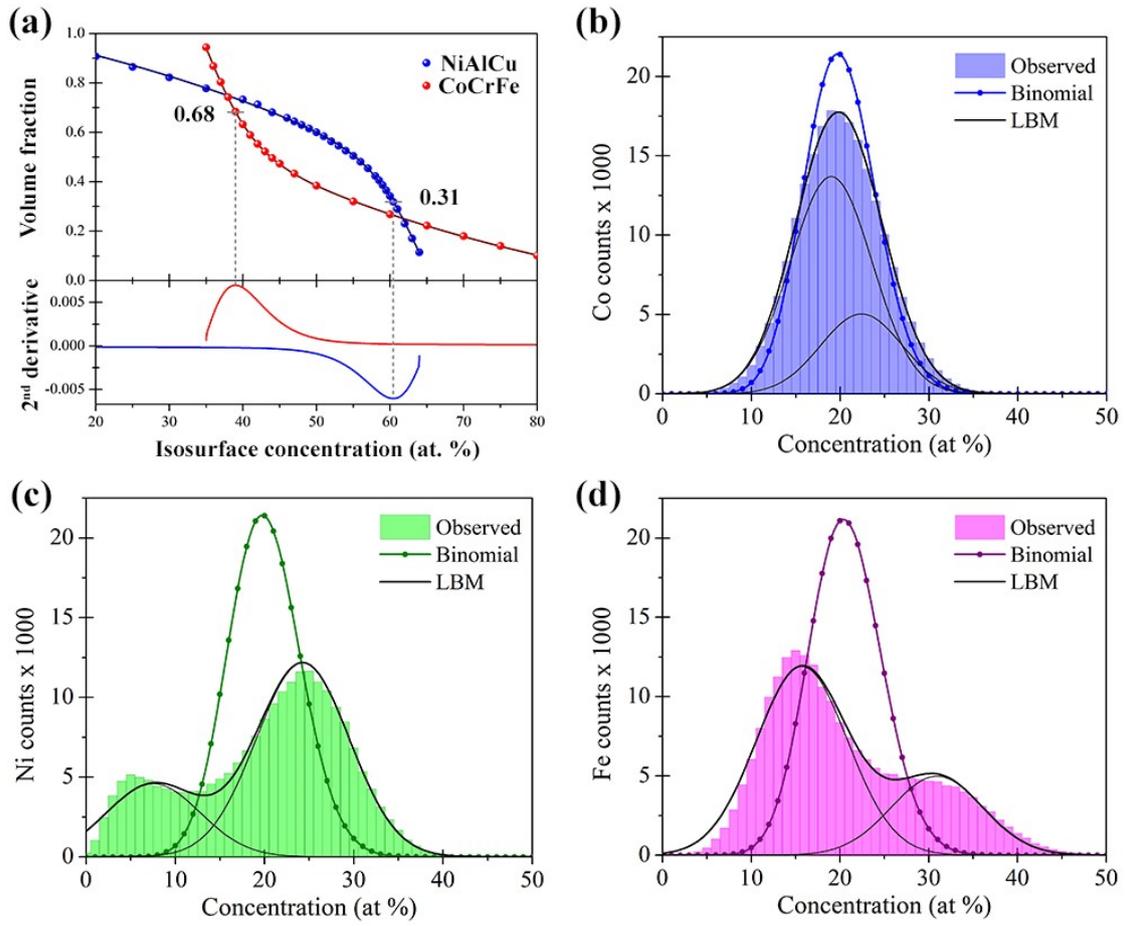

Figure 9: Volume fraction of the domains corresponding to NiAlCu and CoCrFe (a). Frequency analysis for Co (b), Ni (c) and Fe (d) including comparison to binomial and LBM distributions.

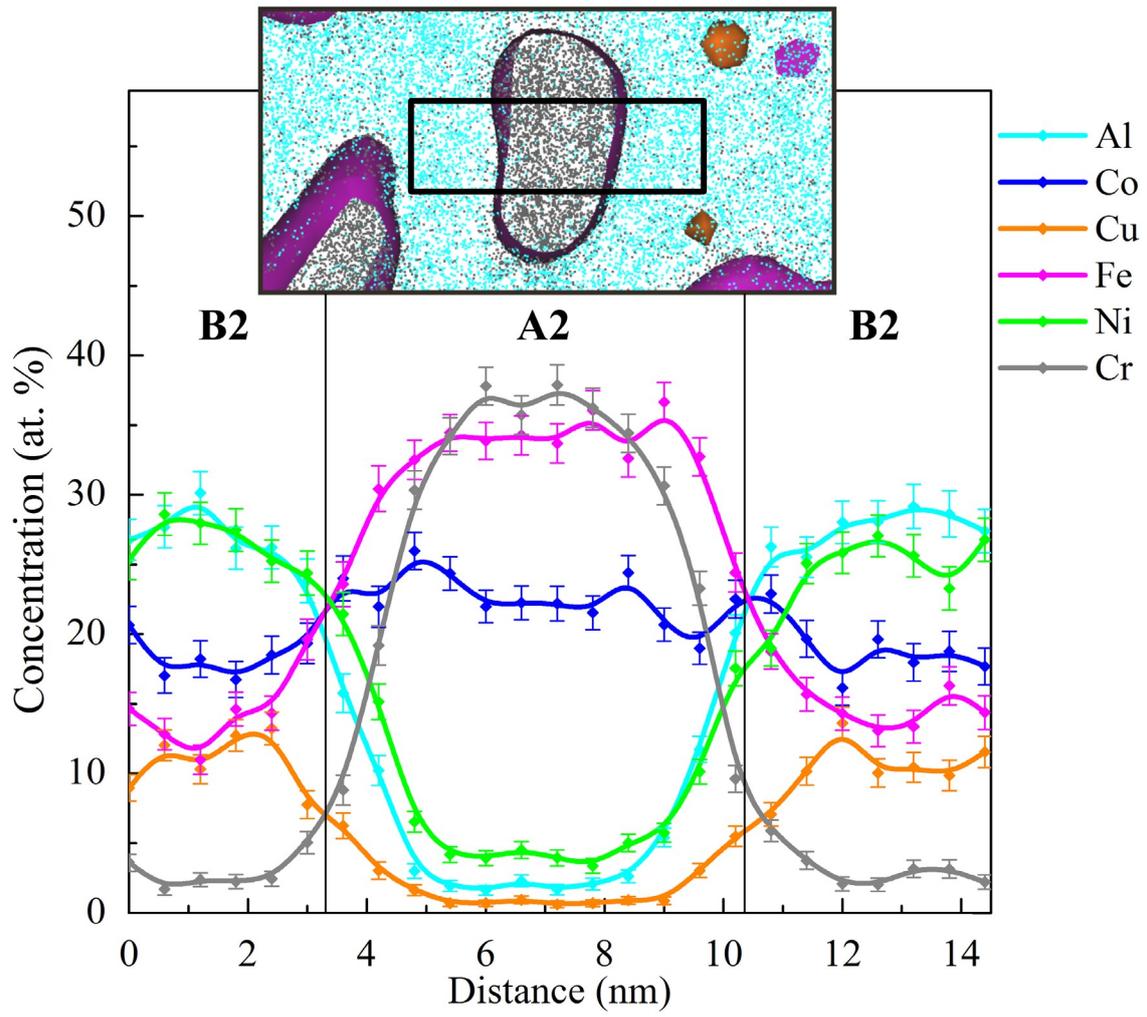

Fig. 10: Concentration profile across a 5 nm Cr-Fe-rich A2 region highlighted in the inset.

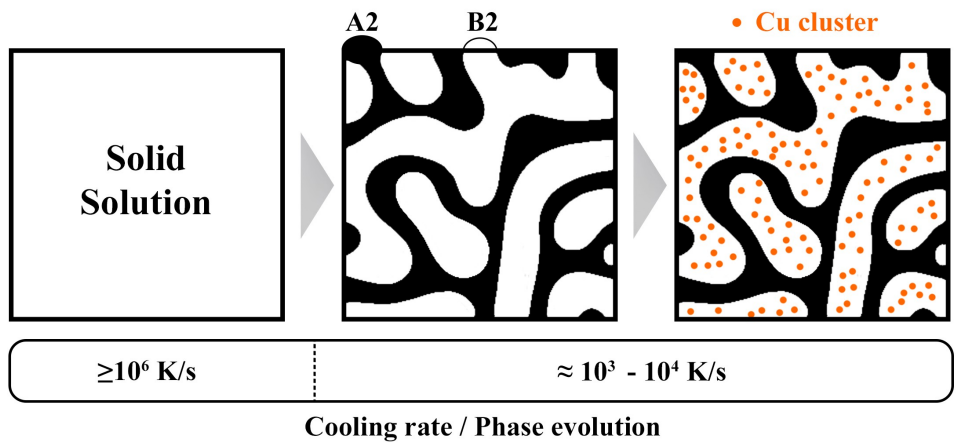

Fig. 11: Schematic illustrations of concentration profiles corresponding to spinodal decomposition and nucleation & growth mechanisms, respectively (a). Illustration of the proposed phase evolution (or phase adoption for certain cooling rates) based on the discussed experimental results, i.e. A2/B2 compartmentation by spinodal decomposition at the nanoscale followed by Cu cluster precipitation through the nucleation & growth mechanism.

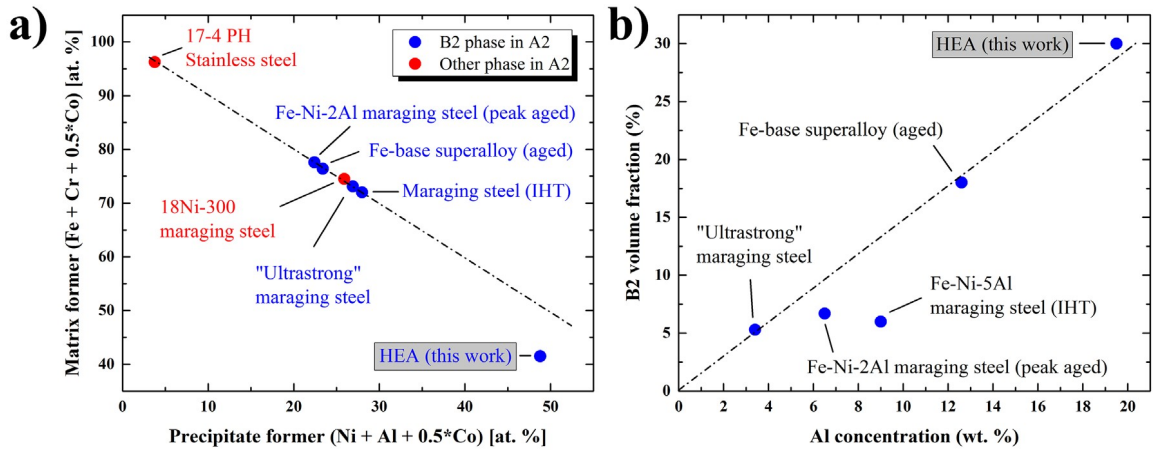

Fig. 12: Effect of precipitate forming elements (Ni and Al) in steel like matrices (Fe and Cr) highlights the analyzed HEA as continuation of a series of Fe-based alloys (mainly steels), containing mostly B2 phases in an A2 matrix (a). The volume fraction of B2 phase in the present alloy can be regarded as continuation of these Fe-based alloys in terms of Al content (b). Data points were taken from [54–57].